\documentclass[aps,prl,twocolumn,preprintnumbers,amsmath,amssymb,colorlinks,showpacs,floatfix]{revtex4}

\usepackage{graphicx}
\usepackage{bm}
\usepackage{soul}
\usepackage{color}
\usepackage{hyperref}

\newcommand {\bacovo} {BaCo$_2$V$_2$O$_8$}

\begin{document}

\title{Longitudinal and transverse Zeeman ladders in the Ising-like chain antiferromagnet \bacovo}

\author{B.~Grenier$^{1,2}$, S.~Petit$^{3}$, V.~Simonet$^{4,5}$, E.~Can\'evet$^{6,7}$, L.-P.~Regnault$^{1,2}$, S.~Raymond$^{1,2}$, 
B.~Canals$^{4,5}$, C.~Berthier$^{8,9}$, P.~Lejay$^{4,5}$}

\affiliation{$^{1}$Univ. Grenoble Alpes, INAC--SPSMS, F-38000 Grenoble, France}
\affiliation{$^{2}$CEA, INAC--SPSMS, F-38000 Grenoble, France}
\affiliation{$^{3}$Laboratoire L\'{e}on Brillouin, CEA--CNRS, CEA-Saclay, F-91191 Gif-sur-Yvette, France}
\affiliation{$^{4}$CNRS, Institut N\'eel, F-38042 Grenoble, France}
\affiliation{$^{5}$Univ. Grenoble Alpes, Institut N\'eel, F-38042 Grenoble, France}
\affiliation{$^{6}$Laboratoire de Physique des Solides, Universit\'{e} Paris-Sud, UMR CNRS 8502, F-91405 Orsay, France}
\affiliation{$^{7}$Institut Laue-Langevin, CS 20156, F-38042 Grenoble Cedex 9, France}
\affiliation{$^{8}$Univ. Grenoble Alpes, Laboratoire National des Champs Magn\'etiques Intenses, F-38000 Grenoble, France}
\affiliation{$^{9}$CNRS, Laboratoire National des Champs Magn\'etiques Intenses, F-38000 Grenoble, France}


\begin{abstract}
We explore the spin dynamics emerging from the N\'eel phase of the chain compound antiferromagnet \bacovo. Our inelastic neutron scattering study reveals unconventional discrete spin excitations, so called Zeeman ladders, understood in terms of spinon confinement, due to the interchain attractive linear potential. These excitations consist in two interlaced series of modes, respectively with transverse and longitudinal polarization. The latter, which correspond to a longitudinal fluctuation of the ordered moment, have no classical counterpart and are related to the zero-point fluctuations that weaken the ordered moment in weakly coupled quantum chains. Our analysis reveals that \bacovo, with moderate Ising anisotropy and sizable interchain interactions, remarkably fulfills the conditions necessary for the observation of discrete long-lived longitudinal excitations.
\end{abstract}

\maketitle


The nature of the excitations in spin half antiferromagnets is a topic of considerable current interest in the field of quantum magnetism. The one-dimensional (1D) case is especially interesting as quantum fluctuations melt the classical long-range N\'eel order. The ground state remains disordered, with a spin excitation spectrum consisting in a continuum composed of pairs of $S=1/2$ excitations called spinons, created or destroyed in pairs, like domain walls in an Ising magnet. Physical realizations of 1D systems, however, eventually order at very low temperature, owing to a small coupling between chains. The metamorphosis of the continuum of spinons that accompany this dimensional cross-over towards a 3D state is an appealing issue \cite{zheludev2002}.

The three possible spin states $S = \pm 1,0$ for a pair of spinons transform upon ordering into two transverse modes and a third collective excitation, which corresponds to fluctuations parallel to the direction of the ordered moment, hence a {\it longitudinal} mode. The observation of the latter is a key issue in condensed matter studies, especially in magnetism with the case of Heisenberg quantum antiferromagnets \cite{Endres2012,Podolsky2011,Merchant2014}, and beyond: If a continuous symmetry is broken, transverse and longitudinal modes are expected in the ordered phase, identified as Goldstone and amplitude modes, respectively \cite{Affleck1992,Schulz1996}. The latter is the analog of what is known in particle theory literature as the Higgs particle \cite{Merchant2014}. Its detection is challenging, since it is generally overdamped because of its decay into massless transverse excitations.

In parallel, another peculiarity of the 1D to 3D dimensional cross-over is that, in the ordered state of quasi-1D systems, each chain experiences an effective staggered molecular field, which gives rise to a linear attractive potential between spinons. The latter competes with their propagating character and finally leads to their confinement in bound states. A spectacular manifestation of this effect in the case of Ising spins, initially described by Shiba \cite{shiba1980}, is the quantization of the transverse excitation continuum in a series of discrete lines below the N\'eel temperature ($T_N$). This effect, called â€œZeeman ladderâ€, was proposed to explain the discretization of the excitations observed in the ordered phase of CsCoCl$_3$ and CsCoBr$_3$ with Raman spectroscopy \cite{ishimura1980,shiba1980}. Recently, a similar series of modes was also observed in the Ising ferromagnetic chain compound CoNb$_2$O$_6$~\cite{coldea2010,morris2014}.

In this article, we introduce a new focus on this physics. We examine the excitations of \bacovo, which realizes an XXZ quasi-1D spin $\frac{1}{2}$ antiferromagnet, intermediate between the Ising and Heisenberg cases. By means of inelastic neutron scattering, we describe below $T_N$ the emergence of {\it long-lived transverse and longitudinal} excitations, in the form of two well defined {\it Zeeman ladders}. The exceptional stability of the longitudinal modes is discussed in connexion with the presence of discretized transverse excitations gapped by the Ising-like anisotropy \cite{Affleck1992,Schulz1996}.

\bacovo\ consists of screw chains of Co$^{2+}$ running along the fourfold ${\bf c}-$axis of the body-centered tetragonal structure \cite{Wichmann1986}. These chains are weakly coupled yielding an antiferromagnetic (AF) ordering (propagation vector ${\bf k}_{AF}=(1,0,0)$ \cite{kimura2008b,canevet2013,SupMat}) in zero field below $T_N \simeq 5.5$~K \cite{He2005,He2006}. The magnetic moment in the distorted octahedral environment is described by a highly anisotropic effective spin $S = 1/2$ \cite{abragam1951} with $g_{xy}=2.95$ and $g_z=6.2$ \cite{kimura2006}, thus allowing quantum fluctuations \cite{kimura2007}. The validity of this description is sustained by the observation of the first crystal field level at 30~meV \cite{SupMat}. This physics is described by the XXZ Hamiltonian:
\begin{equation}
{\cal H } = J \sum_i [\epsilon \left( S_i^x S_{i+1}^x + S_i^y S_{i+1}^y \right)  + S_i^z S_{i+1}^z]
\label{eq1}
\end{equation}
where, according to the analysis of the magnetization curve \cite{kimura2007}, the intrachain AF interaction is $J=5.6$~meV and the anisotropy parameter is $\epsilon=0.46$.


The neutron experiment was performed on the JCNS/CEA--CRG cold neutron three-axis spectrometer IN12 at the Institut Laue-Langevin. A series of energy scans at constant scattering vector ${\bf Q}$ was measured in the N\'eel phase to obtain the spin dispersion parallel and perpendicular to the chains.

\begin{figure}
\begin{center}
\includegraphics[width=8.5 cm]{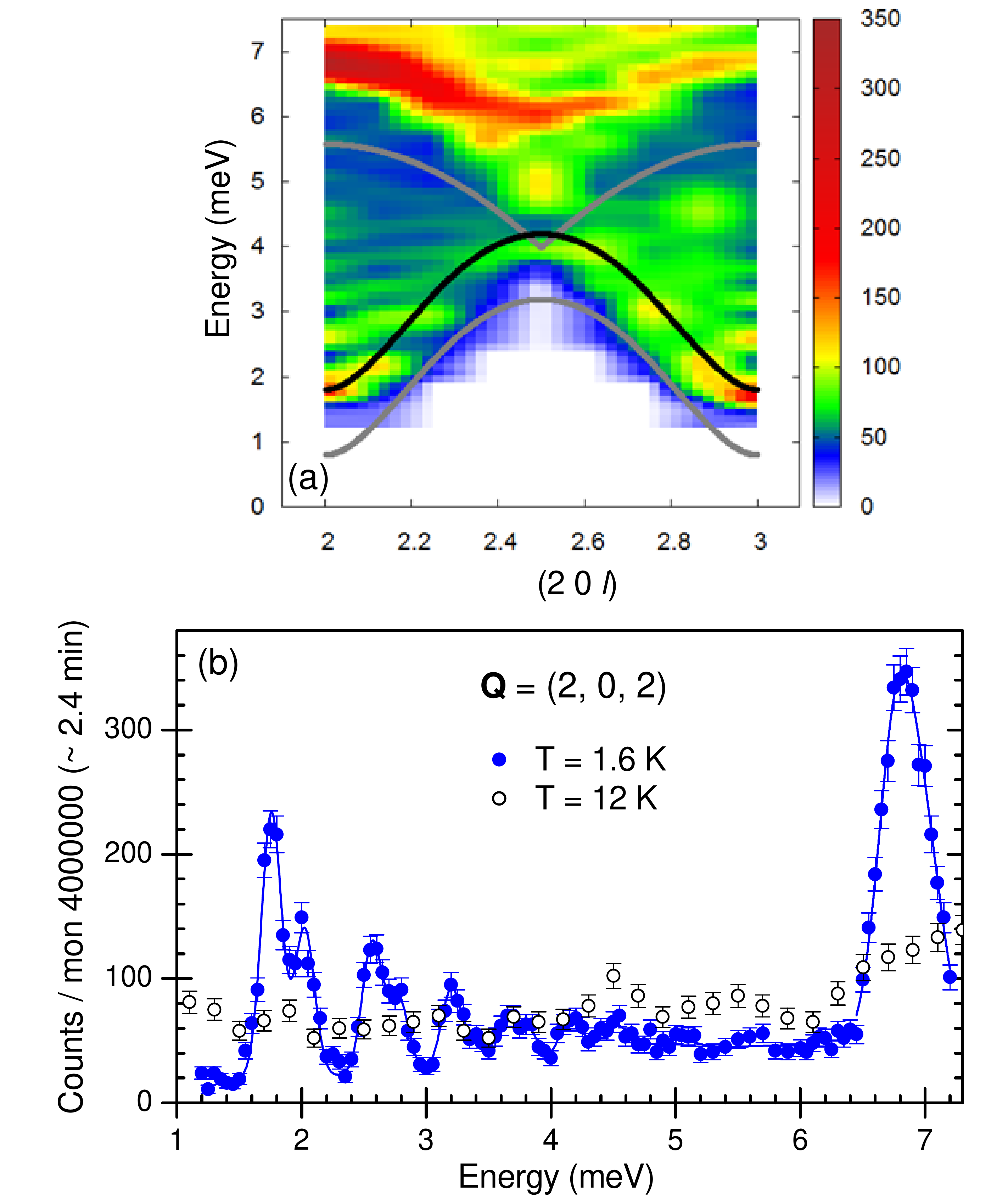}
\caption{\label{Figl}(Color online) (a) Inelastic scattering intensity map obtained from a series of ${\bf Q}$-constant energy scans measured at $T=1.6$~K. The black line is a fit of the lowest mode of the series, $E_1^T$, based on the assumption that its dispersion follows the lower boundary $2E_0^T$ of the two spinon continuum in the purely 1D case (see text). The grey lines indicate the lower and upper boundaries of the corresponding continuum using the fitted $\epsilon = 0.41$ and $J = 2.3$~meV parameters \cite{bougourzi}. (b) Energy scans measured at ${\bf Q} = (2, 0, 2)$ below and above the N\'eel temperature. }
\label{Dispc}
\end{center}
\end{figure}

Direct evidence for the emergence in the ordered phase of unconventional dispersive excitations is shown in Fig.~\ref{Dispc}. At the zone center ${\bf Q} = (2, 0, 2)$, a series of gapped sharp modes ranging between about 1.5 and 6~meV, with decreasing intensities as the energy increases, is observed [see Fig.~\ref{Dispc}(b)]. These 
modes show a sizable dispersion along the chain direction (see Fig.~\ref{Dispc}(a)). The presence of an intense peak
with an out-of-phase weaker dispersion along the ${\bf c}-$axis can also be noticed around 6--7~meV. As expected for magnetic excitations, all these modes disappear above $T_N$ [see Fig.~\ref{Dispc}(b)]. The relative ${\bf Q}$ dependence of their intensities and their energy suggest that the peak around 7~meV can be interpreted as an optical mode, whereas the series of low energy excitations is acoustic-like. The existence of both types of excitations is indeed expected considering the 16 Co$^{2+}$ ions per unit cell in a classical picture. Yet, this intense mode could alternatively be attributed to kinetic bound state of spinons or bound state of pairs of spinons \cite{coldea2010,morris2014}.

\begin{figure} [h]
\begin{center}
\includegraphics[width=8 cm]{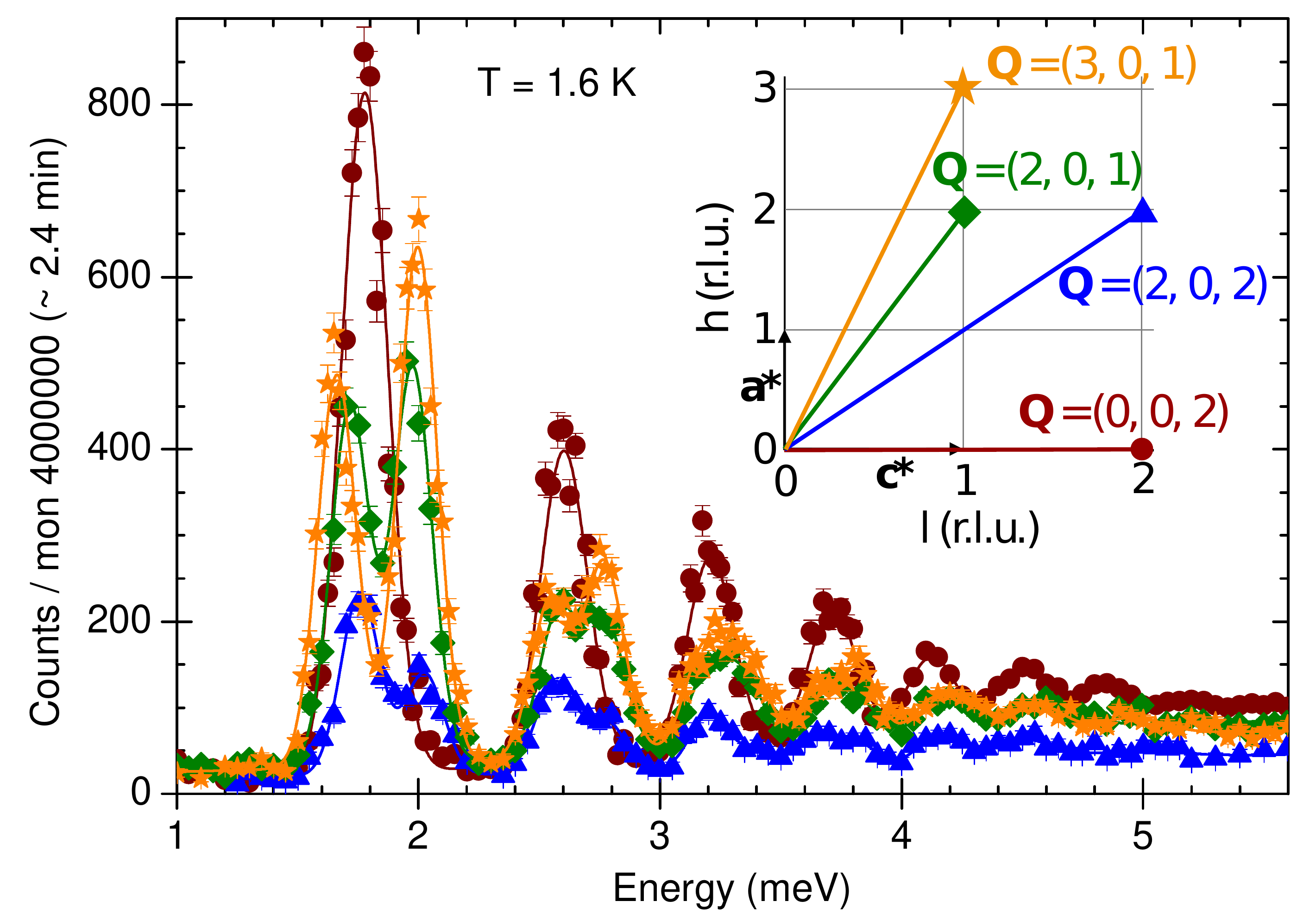}
\caption{(Color online) ${\bf Q}$-constant energy scans, at various Bragg positions shown in the inset (solid symbols), fitted by a series of Gaussian functions (solid lines). This figure emphasizes two series of interlaced sharp $T$ and $L$ modes, the latter arising and increasing in intensity when the ${\bf Q}$-vector rotates from the ${\bf c}-$axis towards the ${\bf a}-$axis direction.}
\label{TL}
\end{center}
\end{figure}

In the following, however, we shall focus on the low energy series, and first investigate their polarization. A neutron scattering experiment is only sensitive to the spin components perpendicular to ${\bf Q}$. Since the ordered moment is along the ${\bf c}-$axis, measurements with ${\bf Q} \parallel {\bf c}$ reveal transverse excitations ($\parallel {\bf a}$ and $\parallel {\bf b}$) while measurements with ${\bf Q} \parallel {\bf a}$ disclose the superposition of transverse ($\parallel {\bf b}$) and longitudinal ($\parallel {\bf c}$) excitations. Energy-scans were thus measured at $T=1.6$~K at various ${\bf Q}$ positions (see Fig.~\ref{TL}). For ${\bf Q} = (0, 0, 2)$, a single series is observed. As the scattering vector rotates towards the ${\bf a}$ direction, a twin series of modes, shifted at slightly higher energies, rises progressively with an intensity that increases with respect to the first series. These results prove unambiguously the transverse ($T$) nature of the first series of discrete modes and
 the longitudinal ($L$) nature of the second one.

\begin{figure}
\begin{center}
\includegraphics[width=7.5cm]{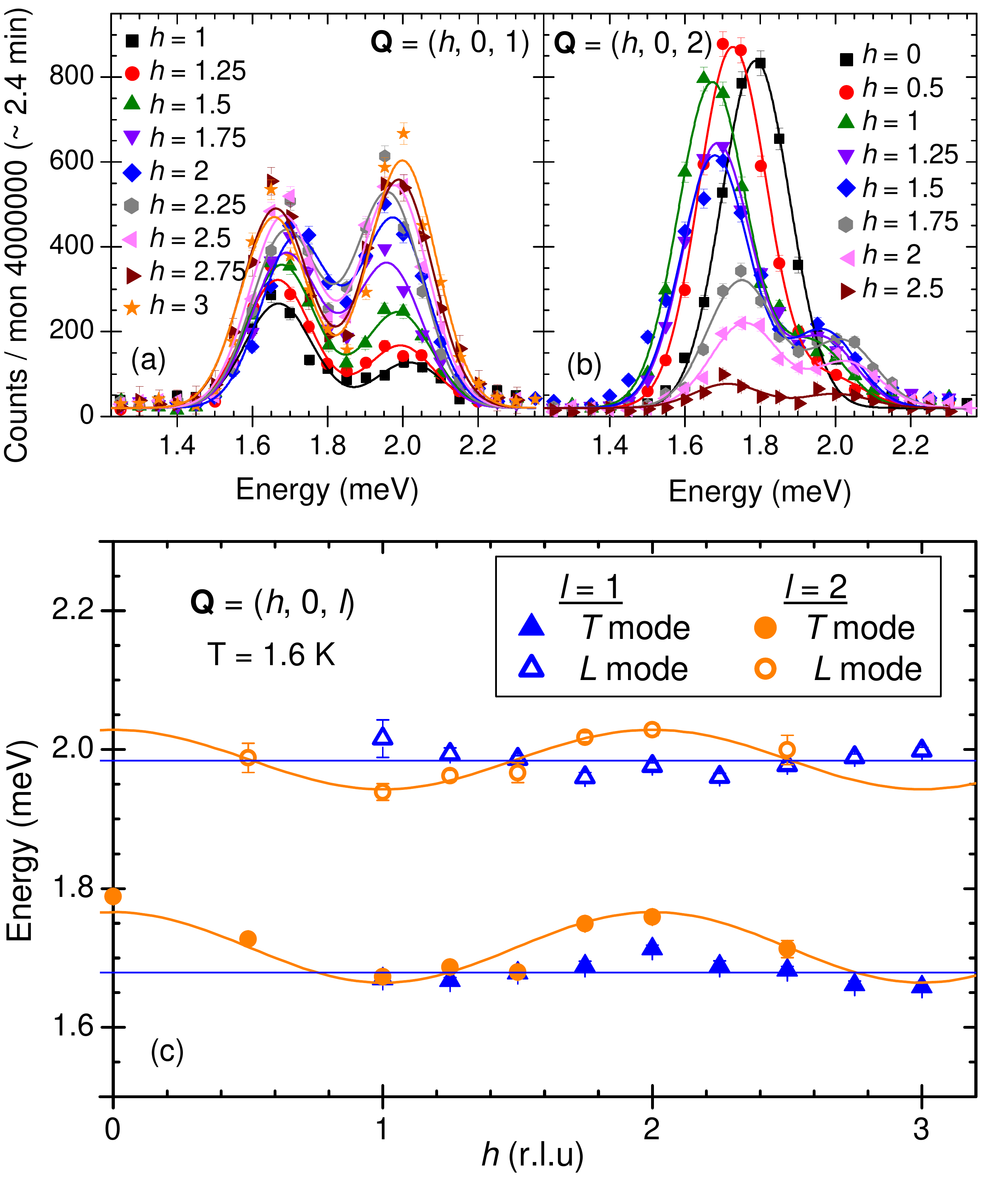}
\caption{\label{Figl} (Color online) ${\bf Q}-$constant energy scans measured at 1.6~K (solid symbols) for (a) ${\bf Q} = (h, 0, 1)$ and (b) ${\bf Q} = (h, 0, 2)$ scanning the lowest $T$ and $L$ modes of the series. These modes are fitted by two Gaussian functions (solid lines), yielding their dispersion curves along ${\bf a^\star}$ plotted in panel (c). The solid lines are guide for the eyes.}
\label{Dispa}
\end{center}
\end{figure}

An important characteristic of this quasi-1D chain system is the strength of the interchain interactions. It can be evidenced from the dispersion of the excitations perpendicular to the chain axis. Figure~\ref{Dispa}(c) presents the energy dependence of the lowest energy mode of the $T$ and $L$ series along ${\bf a^\star}$ obtained from the energy-scans shown in Figs.~\ref{Dispa}(a,b). Although not visible for $l=1$, a sizable dispersion, of the order of 0.1~meV, is observed for $l=2$ with an expected minimum of the gapped mode at the AF points. This peculiar $l$ dependence suggests that the coupling of Co spins belonging to adjacent chains and shifted by $c/2$ should be taken into account, while another exchange interaction in the diagonal direction finally stabilizes the observed magnetic structure \cite{Klanjsek,SupMat}.

\begin{figure}
\begin{center}
\includegraphics[width=7.5cm]{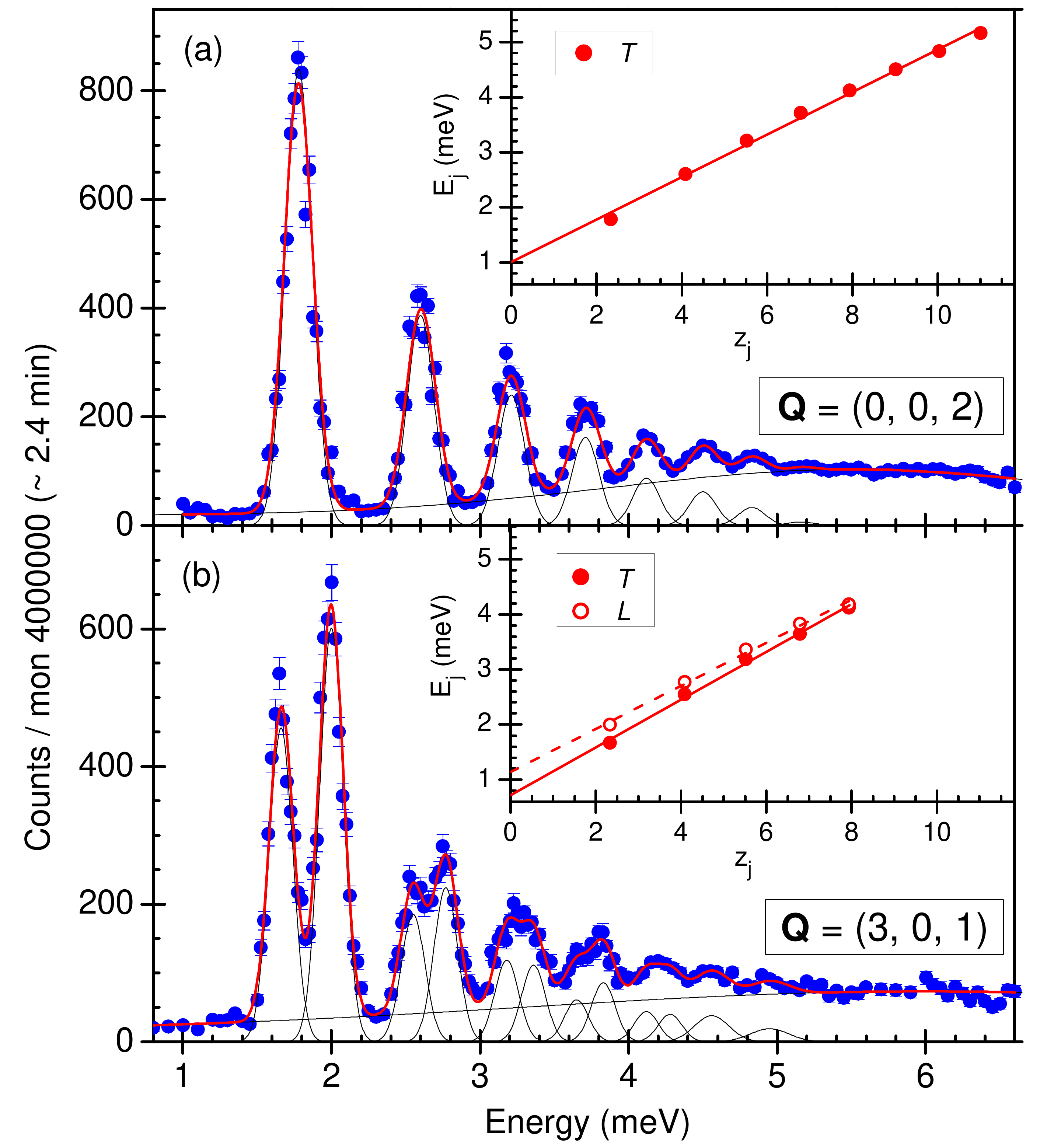}
\caption{\label{Figl} (Color online) Energy scans measured (solid symbols) for (a) ${\bf Q} = (0, 0, 2)$ and (b) ${\bf Q} = (3, 0, 1)$. These series of sharp modes, as well as a broad contribution, were fitted by Gaussian functions (red line for the global fit and black lines for the individual Gaussian functions, see text). The energies of the excitations extracted from the fits are plotted in the insets as a function of the negative zeros $z_j$ of the Airy function (see text). The lines are linear fits to the data.}
\label{Airy}
\end{center}
\end{figure}

Next, we extracted the position of the modes in order to investigate the bounding mechanism of the spinons. The modes at ${\bf Q} = (0, 0, 2)$ and ${\bf Q} = (3, 0, 1)$ were fitted up to 6 meV by a series of Gaussian functions (see Fig.~\ref{Airy}). Their full width at half maximum was obtained from a fit of the lowest energy $T$ and $L$ modes and held constant to the same value (0.2~meV) for the subsequent modes of the series. It was necessary to add an increasing background as the energy increases, 
probably due to a continuum of excitations. For ${\bf Q} = (0, 0, 2)$, eight sharp $T$ modes could be extracted. For ${\bf Q} = (3, 0, 1)$, five $T$ modes and five $L$ modes could be separated. The sixth and seventh modes of the series were fitted by a unique Gaussian function including the $T$ and $L$ modes too close in energy to be separated. This analysis shows that the spacing between the modes appears in a very nontrivial sequencing.

In order to interpret these results, a good starting point is the pure 1D quasi-Ising limit [$\epsilon  \ll 1$ in Eq.~(\ref{eq1})]. A state containing two spinons is created by reversing one or several adjacent spins from one of the 2 degenerate N\'eel states. Two AF bonds are broken, yielding a state with energy $J$, degenerate with all states resulting from reversing an arbitrary number of subsequent spins. These states carry a spin $S_z=\pm1$ for an odd number of reversed spins and $S_z=0$ for an even number. As soon as $\epsilon \ne 0$, the excitation spectrum becomes a continuum composed of such two domain walls which propagate independently. In this picture, the $S_z=\pm1$ states form transverse excitations, while the $S_z=0$ states form longitudinal ones. This 1D domain wall picture and the existence of a gapped continuum were first described by Villain \cite{villain}. Shiba then showed that the introduction of interchain couplings $J'$, acting as a molecular field $h_m$, gives to the two domain wall s
tates an additional potential energy proportional to the distance comprised between them. This causes the above mentioned quantization of the excitation continuum which appears as a series of discrete dispersing lines below the 3D ordering temperature \cite{shiba1980,ishimura1980,coldea2010,kimura2007}.

The linear form of the confining potential imposes that the sequence of excited modes should follow the negative zeros of the Airy function $Ai$~\cite{McCoy1978,car2003,Rutkevich2008,coldea2010}. We thus analyze, at the bound state dispersion minima, the sequence of their energies with:
\begin{equation}
E_j^{T,L} = 2E_0^{T,L} + \alpha~z_j \,\,\,\, {\rm with} \, j=1,2,3,...
\label{eq2}
\end{equation}
with $Ai(-z_j) = 0$ and $\alpha \approx (h_m^2\epsilon J)^{1/3}$ \cite{coldea2010}.
As shown in the insets of Fig.~\ref{Airy}, the energies of the $T$ and $L$ modes were satisfactorily fitted to Eq.~(\ref{eq2}) for various ${\bf Q}$, validating the spinon confinement mechanism. The fit yielded $\alpha \approx 0.42 \pm 0.03$~meV, $2E^{T}_o \approx0.85 \pm 0.15$~meV, and $2E^{L}_o \approx 1.08 \pm 0.05$~meV \cite{SupMat}.
In absence of any microscopic model taking into account both arbitrary $\epsilon$ and interchain interaction (Shiba's model is valid only for $\epsilon  \ll 1$), we then assume that the dispersion along ${\bf c^\star}$ of the first bound state $E_1^T $ is roughly similar to that of the lower boundary of the two-spinon continuum in the pure 1D case, namely $2E_0^T$. For any $J$ and $\epsilon$, this boundary is given by \cite{bougourzi}:
\begin{equation}
2E_0^{T}(l) = \left\{
\begin{array}{ll}
\frac{2I}{1+\kappa} \sqrt{1+\kappa^2-2\kappa \cos{\pi l}} & {\rm for }~ l \le \ell_\kappa\\
\frac{2I}{1+\kappa} \sin{\pi l} & {\rm for }~ \ell_\kappa \le l \le 1-\ell_\kappa\\
\frac{2I}{1+\kappa} \sqrt{1+\kappa^2+2\kappa \cos{\pi l}} & {\rm for }~ l \ge 1-\ell_\kappa
\end{array}
\right.
\label{eq3}
\end{equation}
with $\cos{(\pi \ell_{\kappa})} = \kappa$, $k'=\frac{1-\kappa}{1+\kappa}$, $k=\sqrt{1-k'^2}$, $1/\epsilon = \cosh{(\pi K'/K)}$,  and $J = I \pi/[K\tanh{(\pi K'/K)]}$ ($K$ and $K'$ are the elliptic integrals of argument $k$ and $k'$, respectively) \cite{rq}.

Fitting the dispersion along ${\bf c^*}$ of $E_1^T-\alpha\,z_1=2E_0^T(l)$ with this model in several Brillouin zones gives $J \approx 2.8 \pm 0.4$~meV and $\epsilon \approx 0.41 \pm 0.02$ (see e.g. Fig.~\ref{Dispc}). This analysis locates \bacovo\, in the intermediate anisotropic regime, in agreement with previous estimations of $\epsilon$ \cite{kimura2007,okunishi2007}. Note that $J$ is twice smaller than the estimation given in Refs.~\cite{kimura2006,kimura2007}. Last, 
$h_m$ is directly proportional to an effective interchain interaction $J'_{\mbox{\scriptsize{eff}}}$. 
A quite strong value of {$h_m \sim 0.3$~meV $\propto J'_{\mbox{\scriptsize{eff}}}$ can be inferred from the determination of $\alpha$. This somewhat larger value than the $\sim 0.1$~meV amplitude of the dispersion along ${\bf a^\star}$ is probably due to the frustration between the various interchain couplings \cite{SupMat}.

It is worth noting that in the $\epsilon \ll 1$ limit, the distinguishing feature of the $L$ excitations is the existence of a specific coupling with the N\'eel states. As the $\epsilon$ term exchanges two neighboring spins, the N\'eel state is directly coupled to $S_z=0$ excited states containing 2~reversed spins. This makes the longitudinal modes more massive (at higher energy) than their transverse counterpart, as we observe in \bacovo. The ground state is then an admixture of $S_z =0$ two domain wall states added to the N\'eel state, producing a weakening of the ordered moment. In the $\epsilon \ll 1$ limit, the intensity of the $L$ modes should scale with $\epsilon^2$  \cite{ishimura1980}, explaining why $L$ excitations were hardly observed in systems close to the Ising limit such as CsCoCl$_3$ and CsCoBr$_3$ \cite{shiba1980,Oosawa2006}. The somehow more isotropic character of \bacovo\ (larger $\epsilon$ value) however is expected to enhance the $L$ modes.

In the limit of purely isotropic Heisenberg spins ($\epsilon=1$), a longitudinal mode is also expected \cite{Schulz1996}. It was for instance observed, as a damped excitation, in the antiferromagnetically ordered phases of spin $\frac{1}{2}$ dimer system TlCuCl$_3$ \cite{Merchant2014}, close to the pressure-induced ordering transition, and of the quasi-1D Heisenberg spin $\frac{1}{2}$ antiferromagnet KCuF$_3$ \cite{lake2003}. The longitudinal mode damping is usually attributed to its decay into a pair of gapless transverse spin waves. This longitudinal mode could however not be resolved in another 1D material, namely BaCu$_2$Si$_2$O$_7$, which has a much weaker interchain coupling \cite{zheludev2003}. 
A sufficiently strong dispersion perpendicular to the chains was suggested to be necessary in order to additionally stabilize such a damped longitudinal mode. In \bacovo,  we have indeed determined sizable interchain couplings. Moreover, in contrast to the experimental observation in KCuF$_3$, the \bacovo\ longitudinal modes are remarkably intense and resolution limited. The reason is probably that these longitudinal modes cannot decay into transverse modes since the latter have a large gap, due to the Ising-like anisotropy, and are discretized. It is worth noting that this discretization of longitudinal modes has been reported for the Higgs modes in optical lattice of cold atoms due to confinement \cite{Endres2012}.

It is finally very instructive to recall that \bacovo\ has also raised recently much interest for its field-induced behavior, describable in terms of Tomonaga-Luttinger liquid physics \cite{kimura2008a,kimura2008b,canevet2013}. An exotic magnetic ordered phase, unknown in classical systems, is induced by a magnetic field applied parallel to the chain axis. A longitudinal incommensurate spin density wave (amplitude of the moments modulated along the field direction) is actually stabilized thanks to the particular values of $J'$ and $\epsilon$ \cite{okunishi2007}. Those ingredients, i.e., sizable interchain interactions and intermediate anisotropic character, are the same as the ones we have invoked to account for the quantized transverse and longitudinal magnetic excitations, observed in \bacovo. This material is thus a rare example of spin 1/2 system displaying spin longitudinal modes, of pure quantum origin, in both the dynamical and the field-induced static regimes.

To summarize, our inelastic neutron scattering experiment has revealed unconventional spin excitations in the Ising-like chain antiferromagnet \bacovo: They are quantized due to a weak interchain coupling and consist of two series of both transverse and remarkably strong longitudinal Zeeman ladders. We propose that the stabilization of these longitudinal modes is enabled by the moderate Ising anisotropy prohibiting their decay into the gapped and discretized transverse modes.


We would like to thank R. Ballou, J. Robert, and T. Ziman for fruitful discussions and B. Vettard for his technical support. This work was partly supported by the French ANR project NEMSICOM.


\newpage

\section{Supplemental material}

\subsection{1. Crystalline and magnetic structures of \bacovo}

\bacovo crystallizes in the centrosymmetric tetragonal body-centered $I4_1/acd$ (No. 142) space group, with $a=12.444$~\AA, $c=8.415$~\AA, and eight chemical formulas per unit cell~\cite{Wichmann1986}. The 16 magnetic Co$^{2+}$ ions of the unit cell are equivalent (Wyckoff site $16f$). The spin-3/2 Co$^{2+}$ ions (effective spin-1/2) are arranged in edge-sharing CoO$_6$ octahedra forming screw chains, running along the $c-$axis, and separated by non-magnetic V$^{5+}$ and Ba$^{2+}$ ions (see Fig.~1 in Ref.~\cite{Canevet2013}). Figure~\ref{figureS1} shows one of the two domains of the antiferromagnetic (AF) structure determined in a previous single-crystal neutron diffraction experiment at $H=0$ and $T = 1.8$~K~\cite{Canevet2013}.

\begin{figure} [h]
\begin{center}
\includegraphics[width=6 cm]{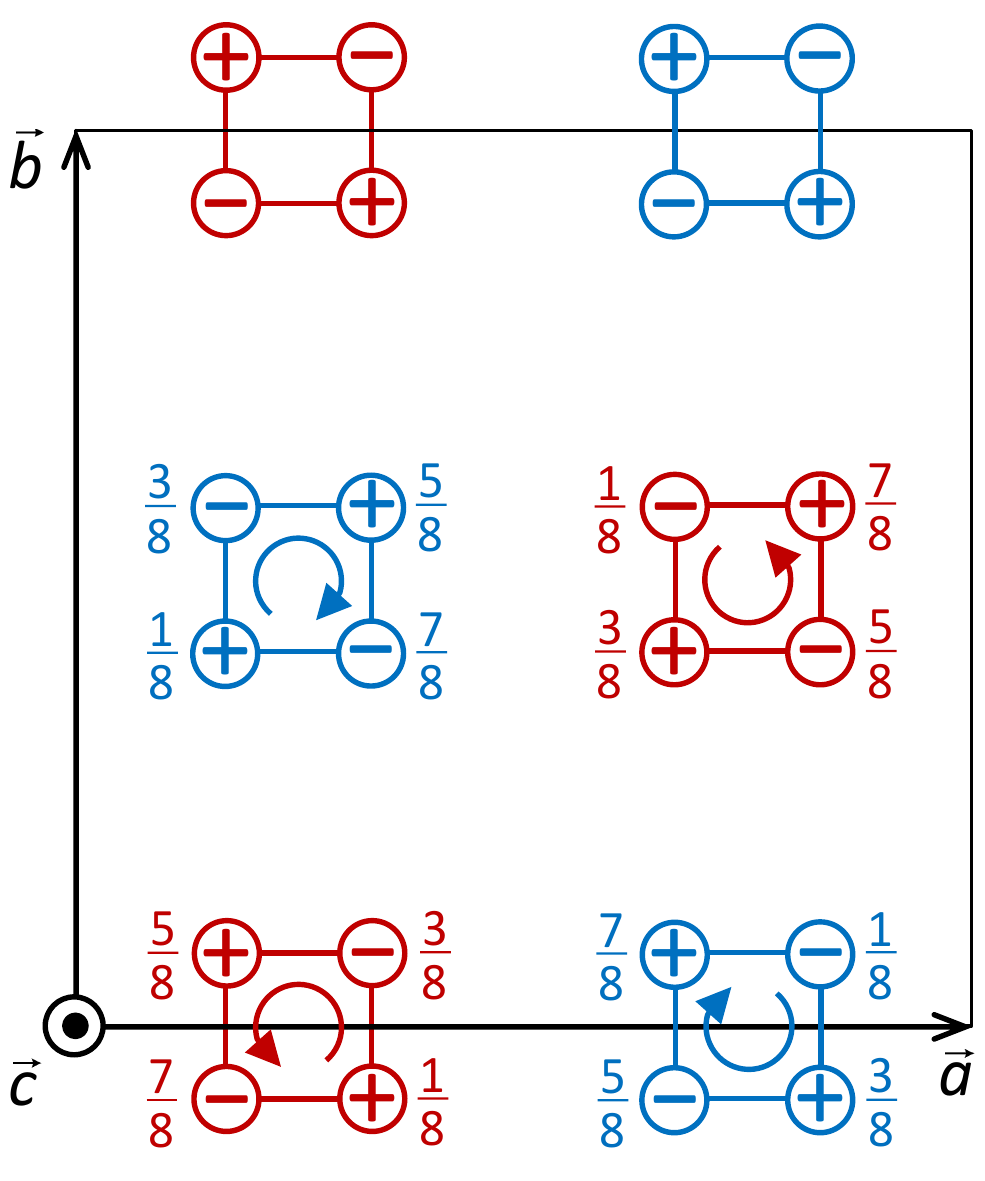}
\caption{(Color online) Magnetic structure in the N\'eel phase of \bacovo determined at $H = 0$ and $T = 1.8$~K~\cite{Canevet2013}. The two types of chains are plotted in projection along the $c-$axis using two different colours: red for the chains described by a $4_1$ screw axis, blue for those described by a $4_3$ axis (the arrows indicate the sense of rotation on increasing $z$). For each Co$^{2+}$ ion of the unit cell, the direction of the spin, '+' or '-', along the $c-$axis is indicated, as well as the $z$ atomic coordinate. This figure presents one of the two magnetic domains; the other domain is simply obtained by reverting all spins in one type of chain, e.g., the blue ones. Notice the 'diagonal' interchain AF coupling between the chains of the same type (e.g., between the 2 Co$^{2+}$ ions located at $z = \frac{3}{8}$ in the two labelled red chains, located at $z = \frac{7}{8}$ in the two blue ones).}
\label{figureS1}
\end{center}
\end{figure}

The dominant interaction is the intrachain nearest neighbor AF exchange coupling (occurring between two Co$^{2+}$ ions of the same chain located at $z=n/8$ and $z=n/8+1/4$, with $n$ integer). This interaction imposes an AF ordering along the chains with the spins parallel to the chain $c-$axis. Looking at the crystalline and AF structures, the dominant interchain interaction is very probably AF along the 'diagonal' direction $a \pm b$, that is between two Co atoms of the same type of chain (blue or red chains) located at the same $z$. This explains the stabilization of the observed two magnetic domains. The various exchange interactions occurring between the two types of chains have been described in details in Ref.~\cite{Klanjsek2012} and were shown to yield an effective 'parallel' (i.e., along the $a$ and $b$ directions) interchain coupling of negligible weight as compared to that of the 'diagonal' interaction.

\subsection{2. Sample and additional neutron scattering data}

The \bacovo single-crystal used in the inelastic neutron scattering (INS) experiments was grown at Institut N\'{e}el (Grenoble, France) by the floating zone method~\cite{Lejay2011}. A 5~cm long cylindrical crystal rod, of about 3~mm diameter, was obtained, with the growth axis at about 60$^\circ$ from the $c-$axis. An about 1 cm thick slice was cut perpendicular to the $c-$axis.

For the neutron experiment performed on the IN12 spectrometer and described in the article, the sample was mounted in a standard cryostat with the $b-$axis vertical. The final wave vector $k_f$ was fixed at 1.5 \AA$^{-1}$ and the higher order contamination was removed using a velocity selector placed before the monochromator.

Additional INS data are presented in Fig.~\ref{CF}. This figure reports measurements obtained on the CEA-CRG thermal neutron three-axis spectrometer IN22 at the Institut Laue-Langevin high-flux reactor, Grenoble, France. The sample was mounted in a standard cryostat with the $b-$axis vertical and the final wave vector $k_f$ was fixed at 3.84 \AA$^{-1}$. Pyrolytic graphite (002) monochromator and analyzer were used, while the $\lambda/2$ contamination was suppressed by using a graphite filter on the incident neutron beam. These measurements show a non-dispersive mode at 30 meV whose intensity decreases with $|Q|$ and which dramatically broadens at high temperature. It is ascribable to the first crystal field level of the Co$^{2+}$ atoms. Note that an alternative explanation of the intense longitudinal modes observed in \bacovo\ could be associated to the true $S=3/2$ nature of the Co$^{2+}$ spin with large anisotropy as described in Ref.~\cite{penc2012}. This explanation is however rather unlikely in view of t
he high energy value of the first crystal field level.

\begin{figure} [h]
\begin{center}
\includegraphics[width = 8 cm]{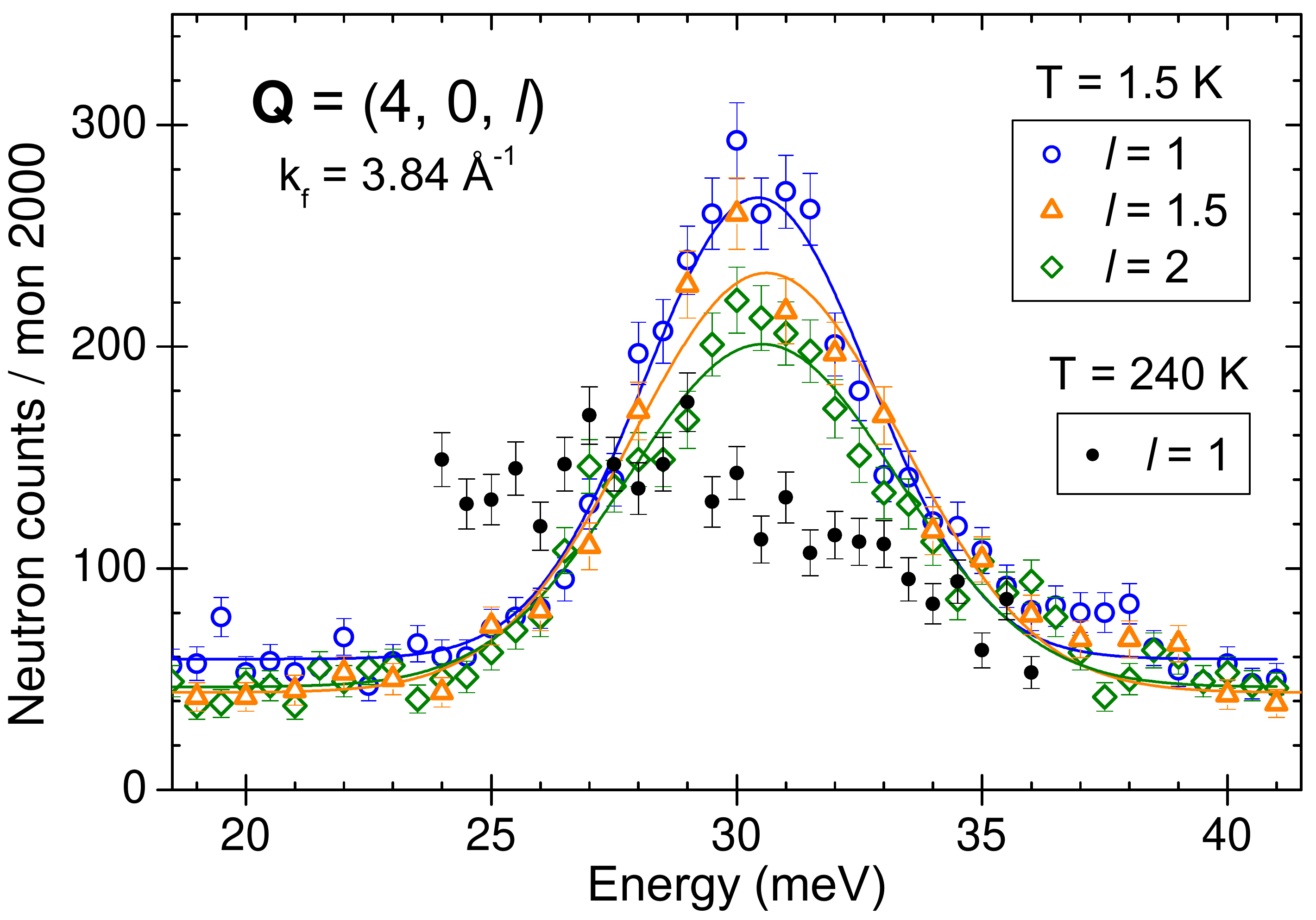}
\caption{(Color online) ${\bf Q}-$constant energy scans (open symbols) measured at various scattering vectors ${\bf Q} = (4, 0, l)$ at $T=1.6$~K and $k_f= 3.84$ \AA$^{-1}$ on IN22, and fitted by a Gaussian function (solid lines). The absence of dispersion for this 30 meV excitation, together with its huge broadening at high temperature (solid black circles), evidence its crystal field nature.}
\label{CF}
\end{center}
\end{figure}

\subsection{3. Additional details about the data analysis}

\begin{table} [h]
\centering
\caption{Threshold energies $2E_0^T$ and $2E_0^L$, coefficient $\alpha$, and agreement factor $r^2$ for the transverse ($T$) and longitudinal ($L$) modes at four different Bragg positions. As the result of the fit slightly depends on the number of modes considered, this number $n_{modes}$ is specified.}
\begin{tabular}{cccccc}
\hline
\hline
${\bf Q}$ ~~ & ~~  $i$ ~~ & ~$2E_0^i$~(meV)~     &   ~$\alpha$~(meV)~  & ~~ $r^2$ ~~ & $n_{modes}$ \\
\hline
$(0, 0, 2)$   &  $T$  &  1.00( 8)   &  0.386(11)     & ~  0.9946 ~ &   8     \\
              &       &  0.85( 8)   &  0.419(13)     &    0.9962   &   5     \\
              &       &  0.79( 6)   &  0.435(12)     &    0.9978   &   4     \\
$(2, 0, 2)$   &  $T$  &  0.79( 9)   &  0.427(18)     &    0.9948   &   4     \\
$(2, 0, 1)$   &  $T$  &  0.77( 8)   &  0.430(15)     &    0.9956   &   5     \\
              &       &  0.72( 9)   &  0.418(13)     &    0.9972   &   4     \\
$(3, 0, 1)$   &  $T$  &  0.71( 9)   &  0.435(16)     &    0.9946   &   5     \\
              &       &  0.66(11)   &  0.448(22)     &    0.9930   &   4     \\
\hline
$(2, 0, 2)$   &  $L$  &  1.10( 8)   &  0.407(16)     &    0.9955   &   4     \\
$(2, 0, 1)$   &  $L$  &  1.08( 7)   &  0.404(12)     &    0.9966   &   5     \\
              &       &  1.03( 7)   &  0.444(18)     &    0.9954   &   4     \\
$(3, 0, 1)$   &  $L$  &  1.13( 9)   &  0.392(15)     &    0.9941   &   5     \\
              &       &  1.06( 6)   &  0.413(11)     &    0.9978   &   4     \\
\hline
\hline
\end{tabular}
\label{TableAiry}
\end{table}

The magnetic Bragg peaks corresponding to the antiferromagnetic structure of \bacovo\ with $k=(0,0,1)$ appear at ${\bf Q}=(h+1,k,l)$ with $h+k+l$ even (condition due to the $I$ type of the lattice). Table~\ref{TableAiry} summarizes, for the scattering vectors shown in Fig.~2, the values of $\alpha$, $2E_0^T$ and $2E_0^L$ obtained by fitting to Eq.~(2) the positions in energy of the transverse (T) and longitudinal (L) discrete modes [see the insets of Figs.~4(a,b) for instance]. The number of modes included in the fits (4 to 8 starting from the lowest energy ones) was varied in order to estimate the error bars. The small dispersion of the results comes from the fact that, as in CoNb$_2$O$_6$ [see Fig. 3(b) in Ref.~\cite{morris2014}],  the energies of the modes do not vary perfectly linearly with the negative zeros of the Airy functions. Note that the threshold energies $2E_0^T$ and 2$E_0^L$, as well as the coefficient $\alpha$, do not depend on the Bragg position. The fitted values, averaged on the various fit
s, are: $2E_0^T=0.85 \pm 0.15$~meV, $2E_0^L=1.08 \pm 0.05$~meV, and $\alpha=0.42 \pm 0.03$~meV.

\newpage

\section{Erratum}

In our recent letter (see pages 1 to 5), we have presented neutron scattering experiments performed on the Ising-like antiferromagnetic chain \bacovo. We showed that in the ordered phase, the spin excitation spectrum consists in two interlaced series of longitudinal and transverse spinon bound states confined by the inter-chain linear potential. 

\begin{figure} [h]
\begin{center}
\vspace{0.4cm}
\includegraphics[width=7.5 cm]{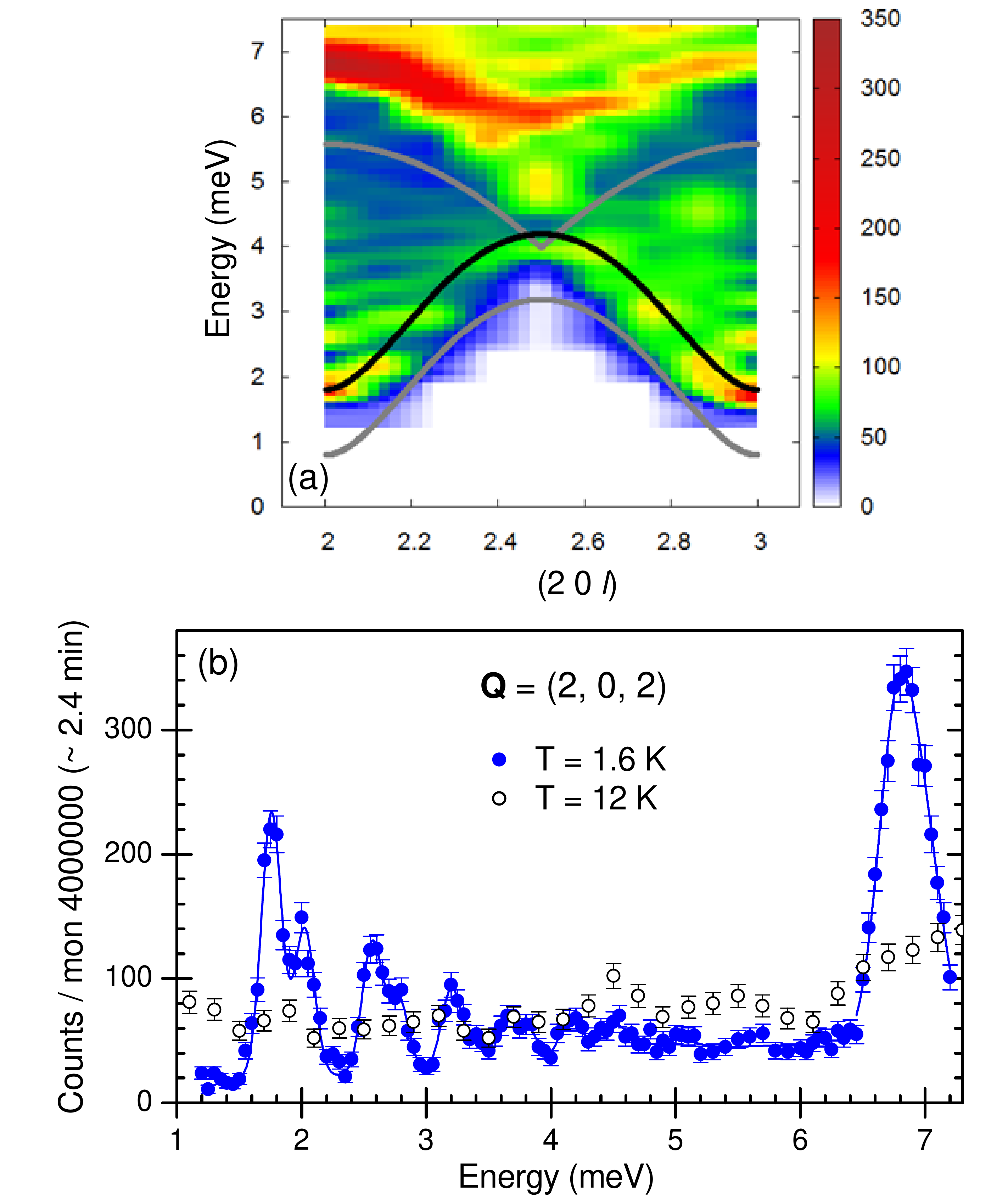}
\caption{\label{Figl}(Color online) Inelastic scattering intensity map obtained from a series of ${\bf Q}$-constant energy scans measured at $T=1.6$~K. The solid black line is the new fit to the lowest mode of the series, $E_1^T$, based on the assumption that its dispersion follows the lower boundary $2E_0^T$ of the two spinon continuum in the purely 1D case. 
The dashed black line comes from the folding of the previous dispersion due to the interchain couplings. Only the lower boundary of the continuum for the pure 1D model is plotted in this figure (grey solid line), the upper one being above the measured energy range. The white dash-dotted lines materialize the splitted anti-crossing branches (phenomenological fit).}
\label{Dispccor}
\end{center}
\end{figure}

To extract the parameters of a model Hamiltonian, we proposed a fit of the lowest mode of the series, $E_1^T$, using the analytical formula given by Bougourzi {\it et al.} (Ref. [22] of our letter). Unfortunately, this fit was made assuming a wrong underlying periodicity of the magnetic structure, neglecting the body-centered type of the unit cell. In this erratum, we correct the fit of the excitation spectrum by taking into account the correct periodicity of the magnetic structure.

Actually, ${\bf Q} = (2, 0, l)$ with $l=2$ corresponds to a zone center (ZC) while that with $l=3$ corresponds to an antiferromagnetic (AF) position. We now impose the minimum and the maximum of the dispersion to be at the AF and ZC positions, respectively, in order to have the correct periodicity (see solid black curve shown in Fig.\,1 of the present erratum). This fit yields larger values for the intrachain coupling, $J=4.8 \pm 0.2$~meV (instead of 2.8~meV) in better agreement with the estimation of Kimura {\it et al.}, $J=5.6$~meV (Refs.~[20,21] of our letter). This also slightly changes the values of the anisotropy parameter, $\epsilon = 0.56 \pm 0.05$ (instead of 0.41) and of the molecular field $h_m$ (thus of the interchain coupling): $h_m \simeq \sqrt{\alpha^3/(\epsilon J)} \simeq 0.2$~meV (instead of 0.3~meV).
 
This interchain coupling causes a folding of the dispersion curve (not present in the case of the 1D model of Bougourzi {\it et al.}), yielding the second dispersion curve plotted with a dashed black line, whose minimum is now at the ZC position. The two branches interact at $l=2.5$ producing an anti-crossing splitting of about 2.2 meV (white dash-dotted curves in Fig.\,1). Another consequence of this new fit is that some spectral weight is now expected up to about 11 meV and that the strong excitation around $6-7$~meV is now well accounted for by the dispersion of the first transverse modes.

To conclude, this corrected model yields a better agreement between the neutron diffraction results and the magnetization measurements of Kimura {\it et al.}. It also gives a better understanding of our complete inelastic scattering intensity map, without affecting the main results and conclusion of our letter. 


\begin{references}
\bibitem{rq} Note that the lower boundary is gapped as soon as $ \epsilon < 1$.
\end{references}
\end{document}